\documentclass[lettersize,journal]{IEEEtran}
\usepackage{amsmath,amsfonts}
\usepackage{algorithmicx,algorithm}

\usepackage{algorithm,algpseudocode}

\usepackage{array}
\usepackage[caption=false,font=normalsize,labelfont=sf,textfont=sf]{subfig}
\usepackage{textcomp}
\usepackage{amssymb}
\usepackage{stfloats}
\usepackage{url}
\usepackage{verbatim}
\usepackage{graphicx}
\usepackage{cite}
\usepackage{bm}      
\begin{document}

\title{Channel-Adaptive Wireless Image Transmission with OFDM}

\author{Haotian Wu,~\IEEEmembership{Graduate Student Member,~IEEE,}
                     Yulin Shao,~\IEEEmembership{Member,~IEEE,}
                     Krystian Mikolajczyk,~\IEEEmembership{Senior Member,~IEEE,}
                     and Deniz G\"{u}nd\"{u}z,~\IEEEmembership{Fellow,~IEEE}
\thanks{The authors are with the Department of Electrical and Electronic Engineering, Imperial College London, London SW7 2AZ, U.K. (e-mail: haotian.wu17@imperial.ac.uk). D. G\"{u}nd\"{u}z is also with the Department of Engineering `Enzo Ferrari', University of Modena and Reggio Emilia (UNI- MORE), Italy.}
\thanks{This work was supported by the European Research Council
(ERC) under Grant 677854, and by the UK EPSRC (EP/W035960/1 and EP/S032398/1) under the CHIST-ERA program (CHIST-ERA-20-SICT-004).
}
}



\maketitle

\begin{abstract}
We present a learning-based channel-adaptive joint source and channel coding (CA-JSCC) scheme for wireless image transmission over multipath fading channels. The proposed method is an end-to-end autoencoder architecture with a dual-attention mechanism employing orthogonal frequency division multiplexing (OFDM) transmission. Unlike the previous works, our approach is adaptive to channel-gain and noise-power variations by exploiting the estimated channel state information (CSI). Specifically, with the proposed dual-attention mechanism, our model can learn to map the features and allocate transmission-power resources judiciously to the available subchannels based on the estimated CSI. Extensive numerical experiments verify that CA-JSCC achieves state-of-the-art performance among existing JSCC schemes. In addition, CA-JSCC is robust to varying channel conditions and can better exploit the limited channel resources by transmitting critical features over better subchannels.
\end{abstract}

\begin{IEEEkeywords} Joint source channel coding, deep neural networks, OFDM, image communications. \end{IEEEkeywords} 

\section{Introduction}
Shannon's separation theorem states that it is optimal to design source and channel codes separately for an infinite block-length \cite{shannon1948mathematical}. However, an increasing number of wireless applications, such as Internet-of-things and edge intelligence \cite{gunduz2020communicate, Jankowski:JSAC:21, shao2021federated}, require the efficient transmission of large volumes of data under strict delay constraints, resulting in an increasing interest in joint source channel coding (JSCC) in recent years. 
Recently, inspired by the success of deep learning techniques, researchers have started to exploit deep neural networks to design novel and competitive JSCC schemes to transmit high information content signals, such as images or videos, over wireless channels \cite{bourtsoulatze2019deep, kurka2020deepjscc, kurka2021bandwidth, yang2021ofdm, choi2019neural, saidutta2021joint, erdemir2021privacy, xu2021wireless, Tung:JSAC22}. This approach has been pioneered in \cite{bourtsoulatze2019deep}, where an autoencoder-based JSCC architecture is proposed for wireless image transmission, which outperformed conventional compression and channel coding schemes over additive white Gaussian noise (AWGN) and Rayleigh fading channels. This was later extended to feedback channels in \cite{kurka2020deepjscc} and to bandwidth-adaptive transmission in \cite{kurka2021bandwidth}. In \cite{yang2021ofdm}, authors consider JSCC over orthogonal frequency division multiplexing (OFDM) channels. An alternative generative architecture is considered in \cite{choi2019neural, saidutta2021joint, erdemir2021privacy}.
\begin{figure}[t] 
    \centering
    \includegraphics[scale=0.36]{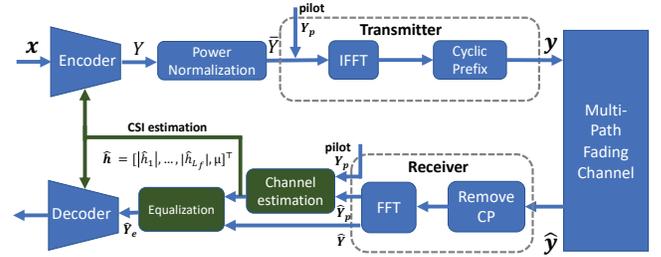}
    \caption{Our proposed channel-adaptive CA-JSCC scheme.}
    \label{OFDM_pipeline}
\end{figure}

\begin{figure*}[t] 
    \centering
    \includegraphics[scale=0.7]{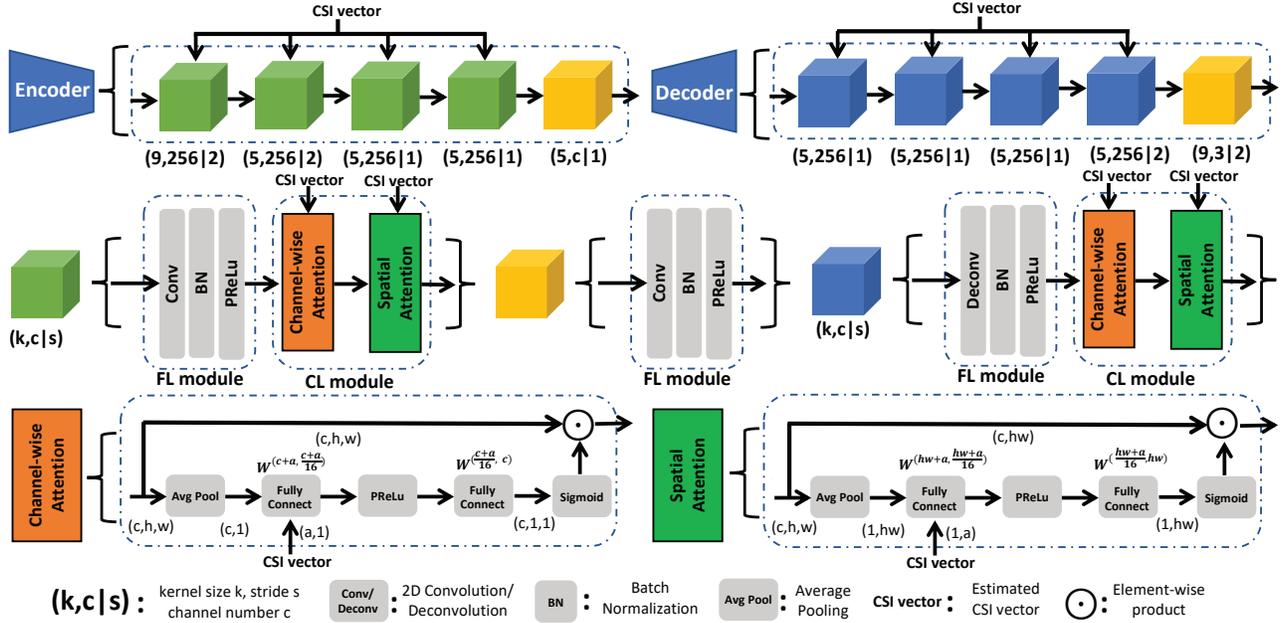}
    \caption{Basic blocks of our dual-attention encoder and decoder architectures.}
    \label{attention_ende}
\end{figure*}

However, adaptability to various channel conditions is still a challenge for deep-learning-based JSCC. Methods in \cite{bourtsoulatze2019deep, kurka2020deepjscc, kurka2021bandwidth} are either trained for a specific signal-to-noise ratio (SNR), or for a range of channel SNRs. The former requires significant storage memory to store different network parameters for different channel conditions, while the latter sacrifices performance and does not exploit the channel state information (CSI). In conventional digital communication systems, CSI at the transmitter can allow power allocation to boost the communication rate. In \cite{xu2021wireless, yang2021ofdm, Tung:JSAC22}, CSI is used in a similar manner in the context of learning-aided design, mainly to adjust the feature weights according to the CSI; however, in the case of OFDM, CSI will be instrumental not only for power control, but also to decide the mapping of the features to different subcarriers according to their relative qualities. 
For example, more critical features of the input image can be mapped to more reliable subcarriers. 

We introduce a channel-adaptive JSCC (CA-JSCC) scheme, which employs a dual-attention mechanism to adjust its features in the multi-scale intermediate layers according to the estimated CSI at the encoder and decoder. Our dual-attention mechanism employs both \textit{channel-wise} attention and \textit{spatial} attention, and jointly learns to map the features to the subcarriers and to allocate power judiciously. Our method achieves state-of-the-art performance and can adapt to different channel conditions.

Our main contributions can be summarized as:

\begin{itemize}
\item To the best of our knowledge, channel adaptability for JSCC with OFDM has not been studied before. All previous methods require the training and testing SNRs to match without fully exploiting the CSI.

\item
We present a CA-JSCC scheme with state-of-the-art performance in various SNR and bandwidth scenarios. We propose a dual-attention mechanism to simultaneously exploit the estimated CSI to aid the allocation of features and power resources to adapt to time-varying channel conditions.

\end{itemize}

\section{System Model}
We consider OFDM-based JSCC of images over a multipath fading channel with $L_t$ paths. We transmit each input image using $N_s$ OFDM symbols accompanied with $N_p$ pilots for channel estimation of $L_f$ OFDM subcarriers. As shown in Fig \ref{OFDM_pipeline}, an encoding function ${E}_{\bm{\theta}}$ first maps the input image $\bm{x}\in \mathbb{R}^{c\times h\times w}$ into a complex matrix $\bm{Y}\in \mathbb{C}^{N_s\times L_{f}}$, where $c,h$ and $w$ denote the color, height and width. The generated channel input can be denoted by $\bm{Y}={E}_{\bm{\theta}}(\bm{x,\hat{h}})$, where $\bm{\hat{h}}$ is the estimated CSI vector available at the transmitter.

Channel input $\bm{Y}$ is subject to an average power constraint $P_s$: $\frac{1}{N_s L_f}\mathbb{E}\big[\|\bm{Y}\|^2_\text{F}\big]\leq P_s$, where the expectation is taken over the input images, and $\|\cdot\|_\text{F}$ denotes the Frobenius norm. Without loss of generality, we set $P_s=1$.

Each OFDM symbol is passed through the inverse discrete Fourier transform (IFFT) module, appended with the cyclic prefix (CP), and transmitted to the receiver over the multipath channel. The transfer function of the multipath fading channel with $L_t$ paths is defined as: $\bm{\hat{y}}=h_c(\bm{y})=\bm{h_t}\ast \bm{y}+\bm{w}$, where $\bm{y}$ and $\bm{\hat{y}}$ denote the input and output vectors, respectively; $\ast$ is the linear convolution operation, $\bm{h_t} \in \mathbb{C}^L$ is the channel impulse response, and $\bm{w}$ is the AWGN term.

The receiver first demodulates $\bm{\hat{y}}$ by removing the CP and  applying fast Fourier transform (FFT). 
The equivalent frequency-domain transfer function of $h_c$ can be written as:
\begin{equation}
\label{fre_eq}
\hat{Y}[i,k]=H[k]\bar{Y}[i,k]+W[i,k],
\end{equation}
where the frequency-domain channel matrix $\bm{H} \in \mathbb{C}^{(L_{f},L_{f})}$ is a diagonal matrix with the $k$-th diagonal element being $H[k]$ (frequency-domain  channel response of the $k$-th subcarrier). Let $\bar{\bm{Y}}\in \mathbb{C}^{(N_s,L_{f})}$ denote the output of the power normalization module at the transmitter (i.e., the inputs to the IFFT module), where $\bar{Y}[i,k]$ denotes the symbol at the $k$-th subcarrier of the $i$-th OFDM symbol. $\bm{W}\in \mathbb{C}^{(N_s,L_{f})}$ is the frequency domain noise matrix, where $W[i,k] \sim \mathcal{CN}(0,\sigma^2)$ and independent of each other.

Given the FFT output of the pilots at the receiver, we use minimum mean square error (MMSE) or least square (LS) channel estimator to estimate the CSI ($H[k]$) in frequency-domain (Eqn. (\ref{fre_eq})). The estimated CSI vector $\bm{\hat{h}}$ is then used to equalize the data (MMSE equalizer). We have
\begin{equation}\label{eq:csi}
\bm{\hat{h}}\triangleq\left[|\hat{h}_1|,|\hat{h}_2|,\cdots,|\hat{h}_{L_f}|,\mu\right]^\top,
\end{equation}
where $\hat{h}_i$, $i=1,...,L_f$, is the estimated channel gain of the $i$-th subcarrier; while $\mu$ is the average SNR defined as $\mu=10\log_{10} \frac{P_{s}}{\sigma^2}$ dB for transmit power $P_s$ and noise power $\sigma^2$.

Based on the equalized data $\bm{\hat{Y}_e}$ and the estimated CSI vector $\bm{\hat{h}}$, the decoder $D_{\bm{\phi}}$ reconstructs the transmitted image as $\bm{\hat{x}}$, i.e., $\bm{\hat{x}}=D_{\bm{\phi}}(\bm{\hat{Y}_e},\bm{\hat{h}})$. The performance indicator is the peak signal-to-noise ratio (PSNR), defined as $\text{PSNR}=10\log_{10}\frac{(\max{\bm{x}})^2}{\text{MSE}(\bm{x},\bm{\hat{x}})}~(\text{dB})$, where $\max {\bm{x}}$ denotes the maximum possible value of the input signal $\bm{x}$, $\text{MSE}(\bm{x},\bm{\hat{x}})\triangleq E[\|\bm{x}-\bm{\hat{x}}\|^2_2]$ and the expectation is taken over all pixels.

We then jointly train the encoder and decoder to minimize the loss function $\mathcal{L}(\bm{\theta},\bm{\phi} )=\mathbb{E}\big[\text{PSNR}(\bm{x},\bm{\hat{x}})\big]$, where the expectation is taken over the randomness both in the source and channel distributions. 

\begin{figure*}[tb]
    \centering
    \subfloat[]{
    \label{PA_all_SNR} 
    \begin{minipage}[t]{0.32\linewidth}
    \centering
    \includegraphics[scale=0.33]{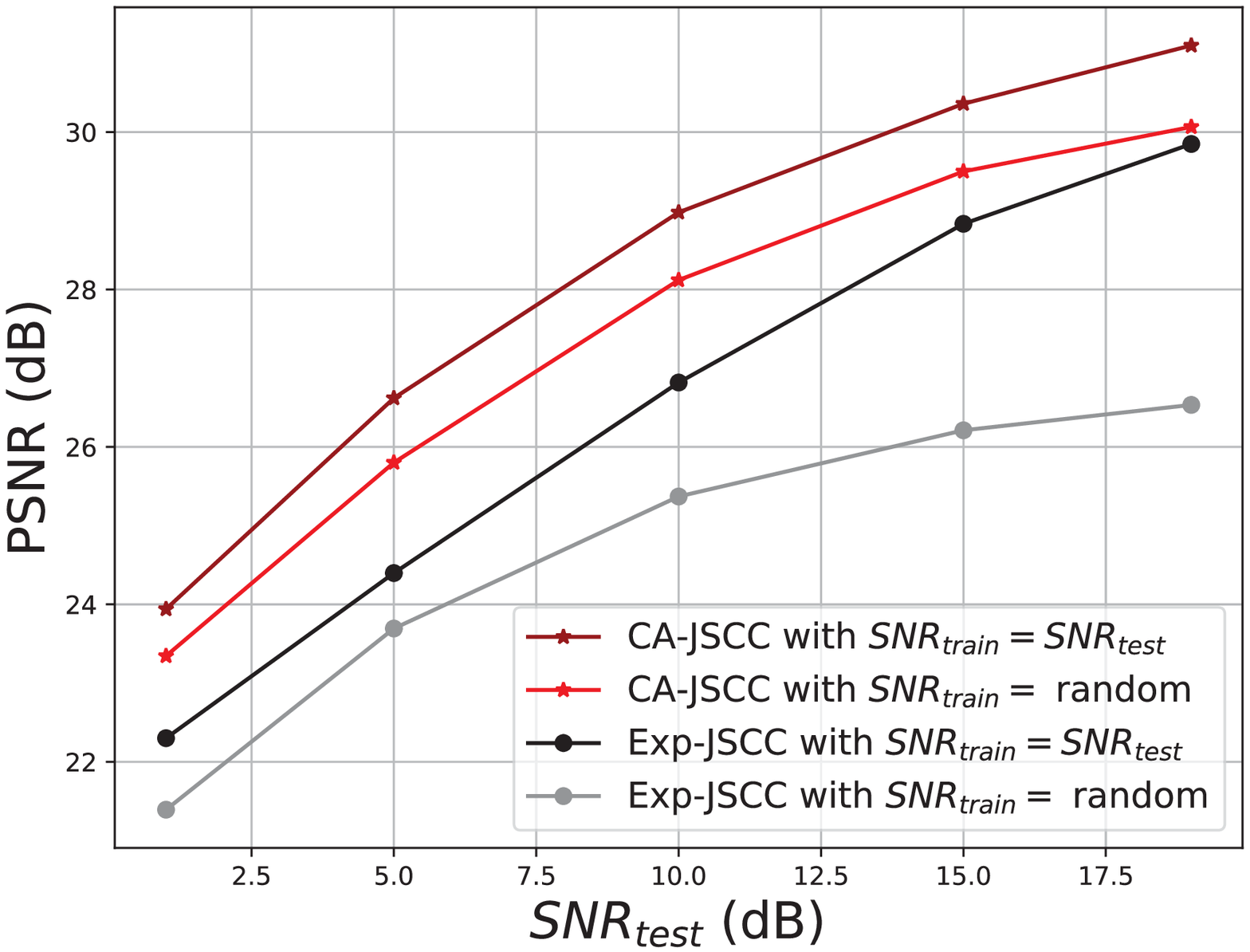}
    \end{minipage}%
    }%
    \subfloat[]{
    \label{Ab_attention} 
    \begin{minipage}[t]{0.32\linewidth}
    \centering
    \includegraphics[scale=0.33]{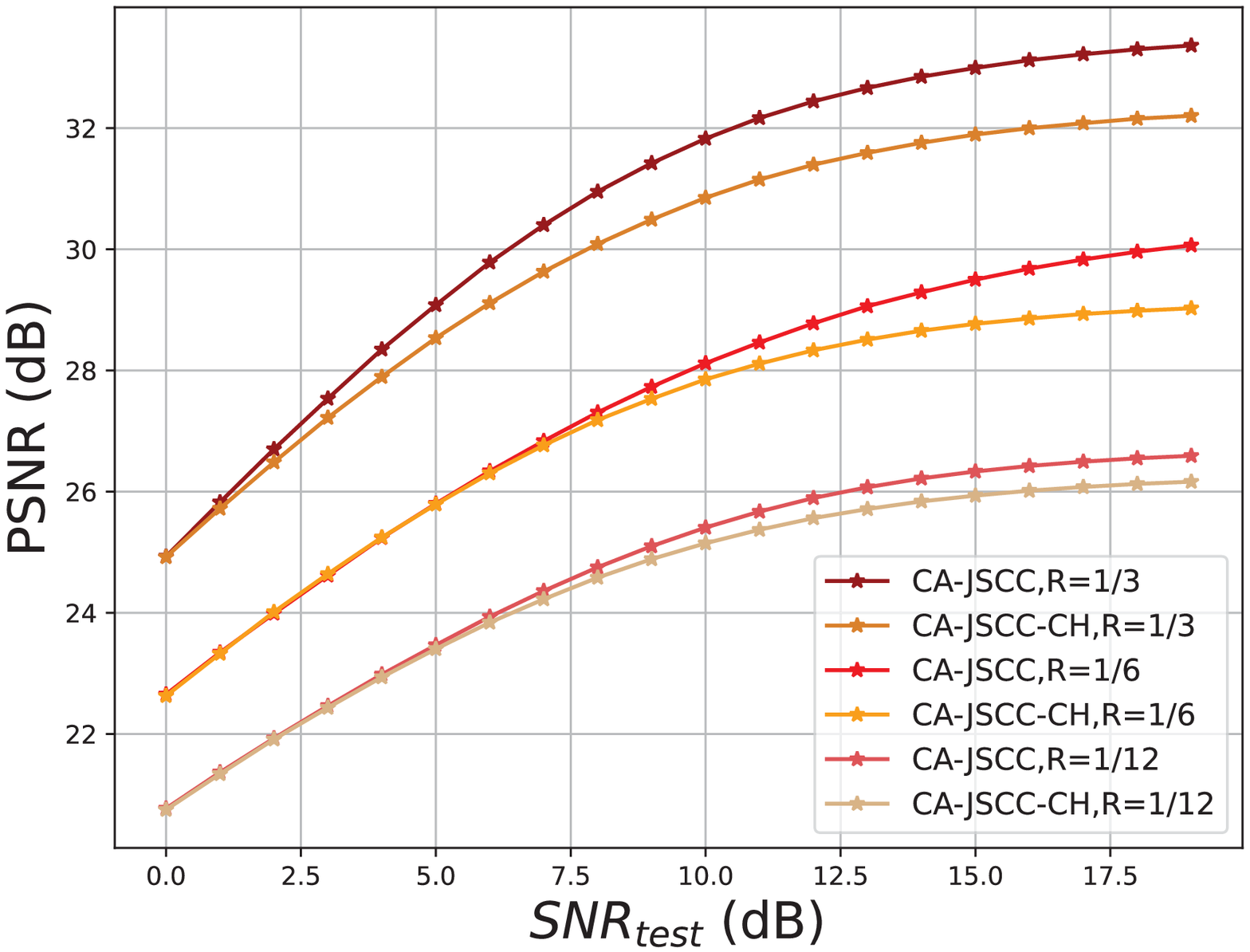}
    \end{minipage}%
    }%
    \subfloat[]{
    \label{power_allocation} 
    \begin{minipage}[t]{0.32\linewidth}
    \centering
    \includegraphics[scale=0.33]{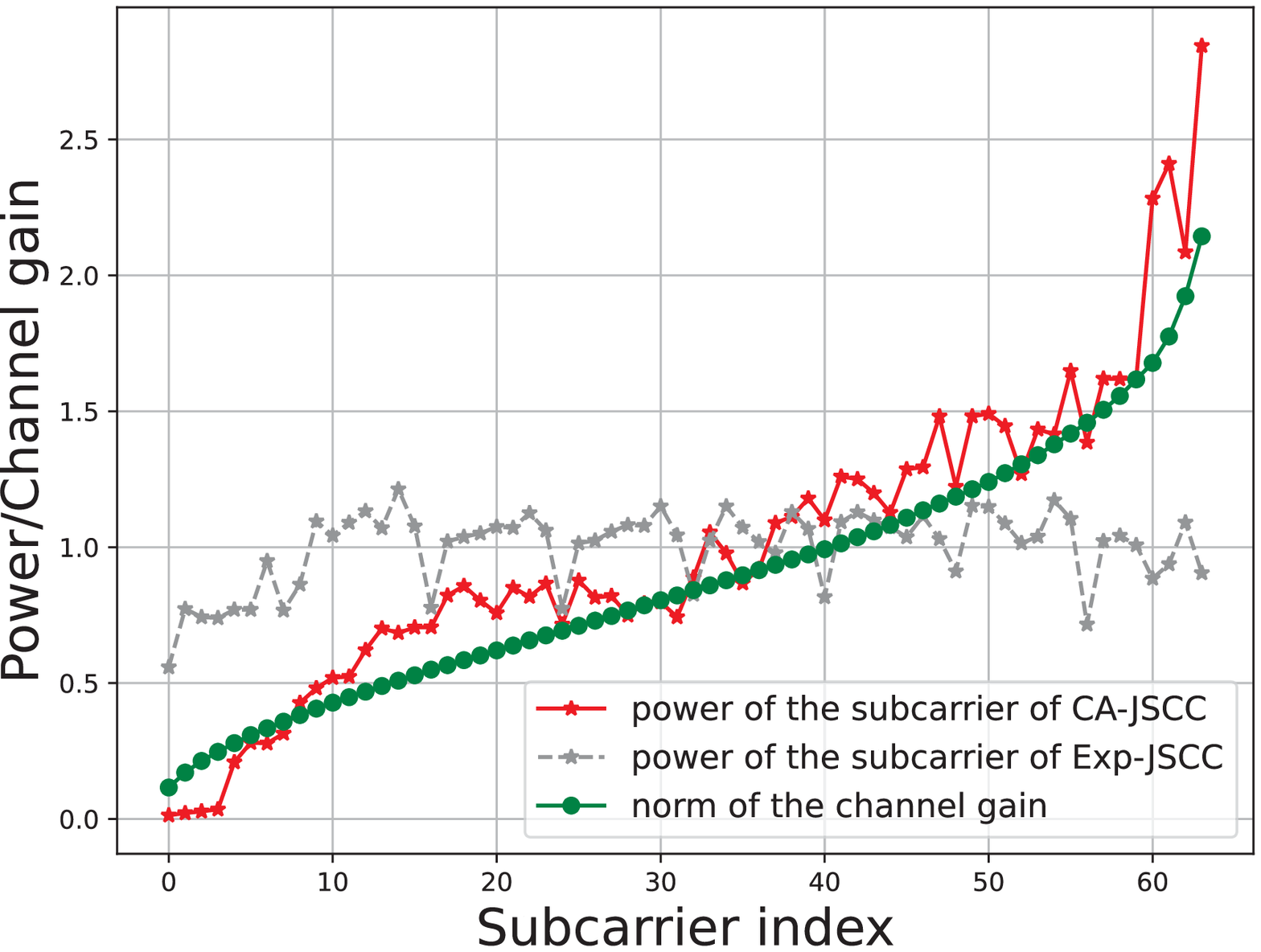}
    \end{minipage}%
    }%
    \centering
    \caption{(a) Comparison of our CA-JSCC scheme with Exp-JSCC when the values of the train and test SNRs are the same. (b) Ablation experiments for the attention strategy. (c) Visualization of the power allocation for the CA-JSCC and the Exp-JSCC.}
\end{figure*}

\section{Dual-attention mechanism}
In OFDM systems, different subcarriers face different channel gains, and a judicious transmission scheme should be able to allocate power and features across appropriate subcarriers to adapt to channel variations. To fully exploit the estimated CSI, we propose a dual-attention based CA-JSCC scheme.

The architecture of our method is shown in Fig \ref{attention_ende}, both the encoder and the decoder have five feature learning (FL) and four channel learning (CL) modules in a comb structure, allowing modulating each feature with a different scale. The FL module consisting of 2D convolution/deconvolution, batch normalization, and PReLU layers, is designed to learn an intermediate representation of the input. The dual-attention based CL module is designed to learn an attention mask to map the intermediate features to appropriate subcarriers based on the estimated CSI and input features. The CL module consists of a \textit{channel-wise} attention module and a \textit{spatial} attention module. Its operation is presented in Algorithm \ref{two_stage} in detail.

\begin{algorithm}[t] 
    \caption{Dual-attention based CL}
    \textbf{Input:} $\bm{F_{in}}\in \mathbb{R}^{c\times h\times w}$, $\bm{\hat{h}}\in \mathbb{R}^{L_{f}+1}$\\
    \textbf{Output:} $\bm{F_{out}}\in \mathbb{R}^{c\times h\times w}$\\
    \textbf{Stage 1: Channel-wise attention} \\
    \vspace{-15pt}
    \begin{algorithmic}[1]
    \State \texttt{$\bm{F_{ca}}=Ave_c(\bm{F_{in}}) \in \mathbb{R}^{(c)}$}
    \State \texttt{$\bm{i_c}=concatenate(\bm{F_{ca}},\bm{\hat{h}})\in \mathbb{R}^{(c+L_{f}+1)}$}
    \State \texttt{$\bm{S_c}=f_c(\bm{i_c}) \in \mathbb{R}^{(c,1,1)}$}
    \For{i= 0:1:c} 
    \State \texttt{$\bm{F_{cout}}[i,h,w]=\bm{S_c}[i] \odot \bm{F_{in}}[i,h,w] \in \mathbb{R}^{(c,h,w)}$}
     \EndFor
    \end{algorithmic}
    \textbf{Stage 2: Spatial attention} \\
    \vspace{-15pt}
    \begin{algorithmic}[1]
    \State \texttt{$\bm{F_{sa}}=Ave_s(\bm{F_{cout}}) \in \mathbb{R}^{(1,h,w)}\Rightarrow \mathbb{R}^{(hw)}$}
    \State \texttt{$\bm{i_s}=concatenate(\bm{F_{sa}},\bm{\hat{h}})\in \mathbb{R}^{(hw+L_{f}+1)}$}
    \State \texttt{$\bm{S_s}=f_s(\bm{i_s}) \in \mathbb{R}^{(hw)}\Rightarrow \mathbb{R}^{(h,w)}$}
    \For{j= 0:1:h}
    \For{k= 0:1:w}
    \State \texttt{$\bm{F_{out}}[c,j,k]=\bm{S_s}[j,k] \odot \bm{F_{cout}}[c,j,k]\in \mathbb{R}^{(c,h,w)}$}
     \EndFor
     \EndFor
    \end{algorithmic}
    \label{two_stage}
    \end{algorithm}

\subsubsection{Channel-wise attention module}
Our \textit{channel-wise} attention module is inspired by\cite{xu2021wireless}, which adapts to a single SNR value in an AWGN channel model. Instead, CA-JSCC learns an attention mask to allocate and map features based on the estimated CSI of all $N_s$ subcarriers.

We first apply average pooling $Ave_c(\bm{F_{in}})$ on input features $\bm{F_{in}}$ along the spatial direction to get vector $\bm{F_{ca}}$, where $Ave_c(\bm{F_{in}})\triangleq \frac{1}{hw}\sum_{j=1}^h\sum_{k=1}^wF_{in}[c,h,w]$. $\bm{F_{ca}}$ is then concatenated with $\bm{\hat{h}}$ to get the intermediate vector $\bm{i_c}$ to compute the \textit{channel-wise} attention mask $\bm{S_c}$ by several fully connected (FC) layers: $S_c=f_c(\bm{i_c})$, where $f_c$ represents the FC layers followed by PReLU functions. Finally, we get the output of our \textit{channel-wise} attention module as $\bm{F_{cout}}=\bm{S_c} \odot \bm{F_{in}}$.

Our \textit{channel-wise} attention module learns to map features from the input to the subcarriers based on the estimated CSI, allowing JSCC to dynamically adjust to different channel SNRs. But the spatial information is ignored when computing the channel attention mask by the average pooling operation. So we design a \textit{spatial} attention module to compensate for spatial information.

\subsubsection{Spatial attention module}
Our \textit{spatial} attention module learns to match the more critical spatial features along the $h$ and $w$ dimensions with better channel conditions depending on the estimated CSI.

This time we firstly apply average pooling $Ave_s(\bm{F_{cout}})$ on the $\bm{F_{cout}}$ along the channel direction to get $\bm{F_{sa}}$, where  $Ave_s(\bm{F_{cout}})\triangleq \frac{1}{c}\sum_{j=1}^cF_{cout}[c,h,w]$. Then, $\bm{\hat{h}}$ and $\bm{F_{sa}}$ are concatenated to get the intermediate feature $\bm{i_s}$, which is used to compute the \textit{spatial} mask $\bm{S_s}$ by several FC layers. We compute the final output feature vector as $\bm{F_{out}}=\bm{S_s} \odot \bm{F_{cout}}$.

The \textit{spatial} attention module can further improve the PSNR performance by exploiting the spatial information and helping JSCC encoder to do more adaptive power allocation, which matches critical features with better channels. 

\begin{figure*}[tb]
    \centering
    \subfloat[$R=1/12$]{
    \label{attention_JSCC_12} 
    \begin{minipage}[t]{0.32\linewidth}
    \centering
    \includegraphics[scale=0.33]{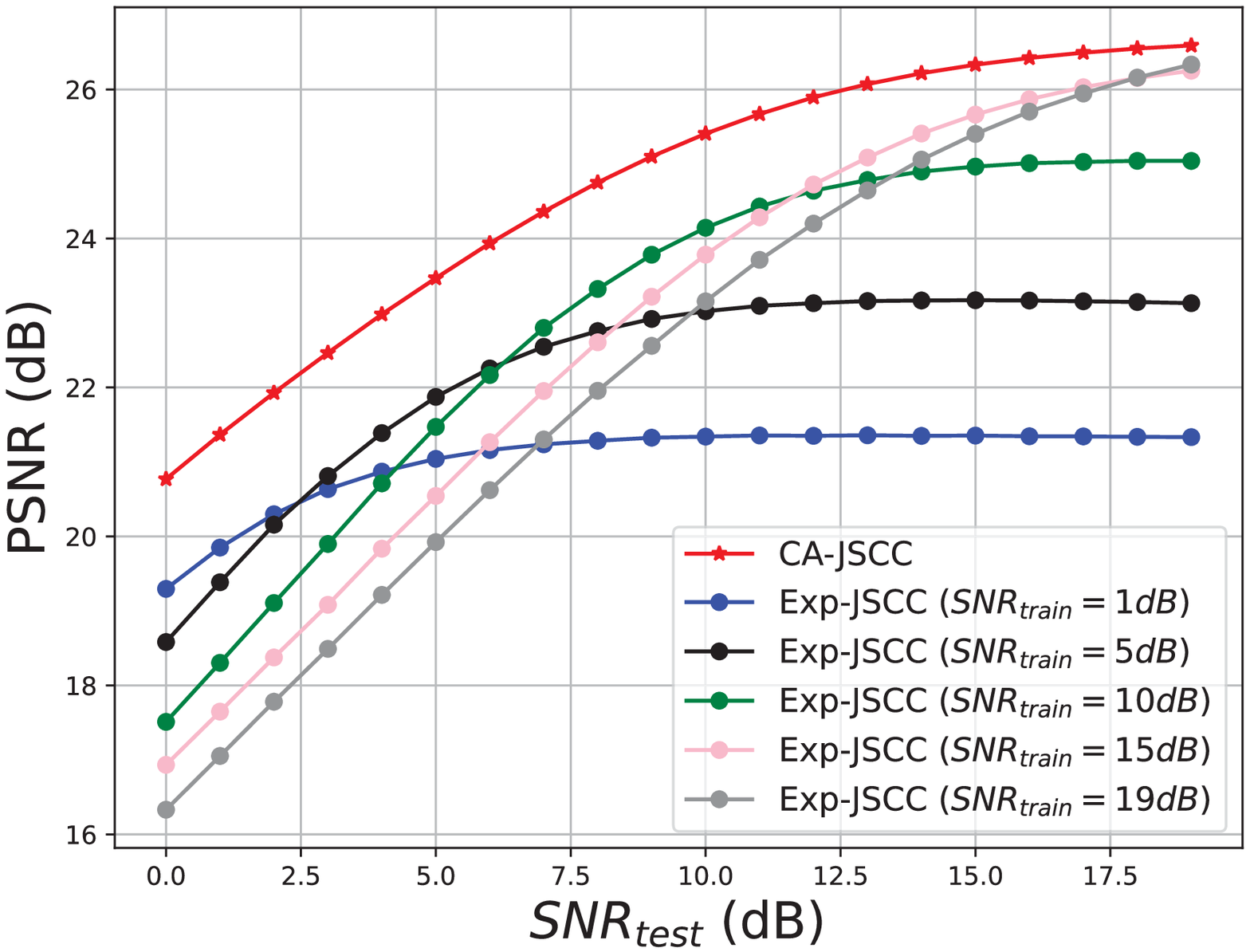}
    \end{minipage}%
    }%
    \hspace{-0.5mm}
    \subfloat[$R=1/6$]{
    \label{attention_JSCC_6} 
    \begin{minipage}[t]{0.32\linewidth}
    \centering
    \includegraphics[scale=0.33]{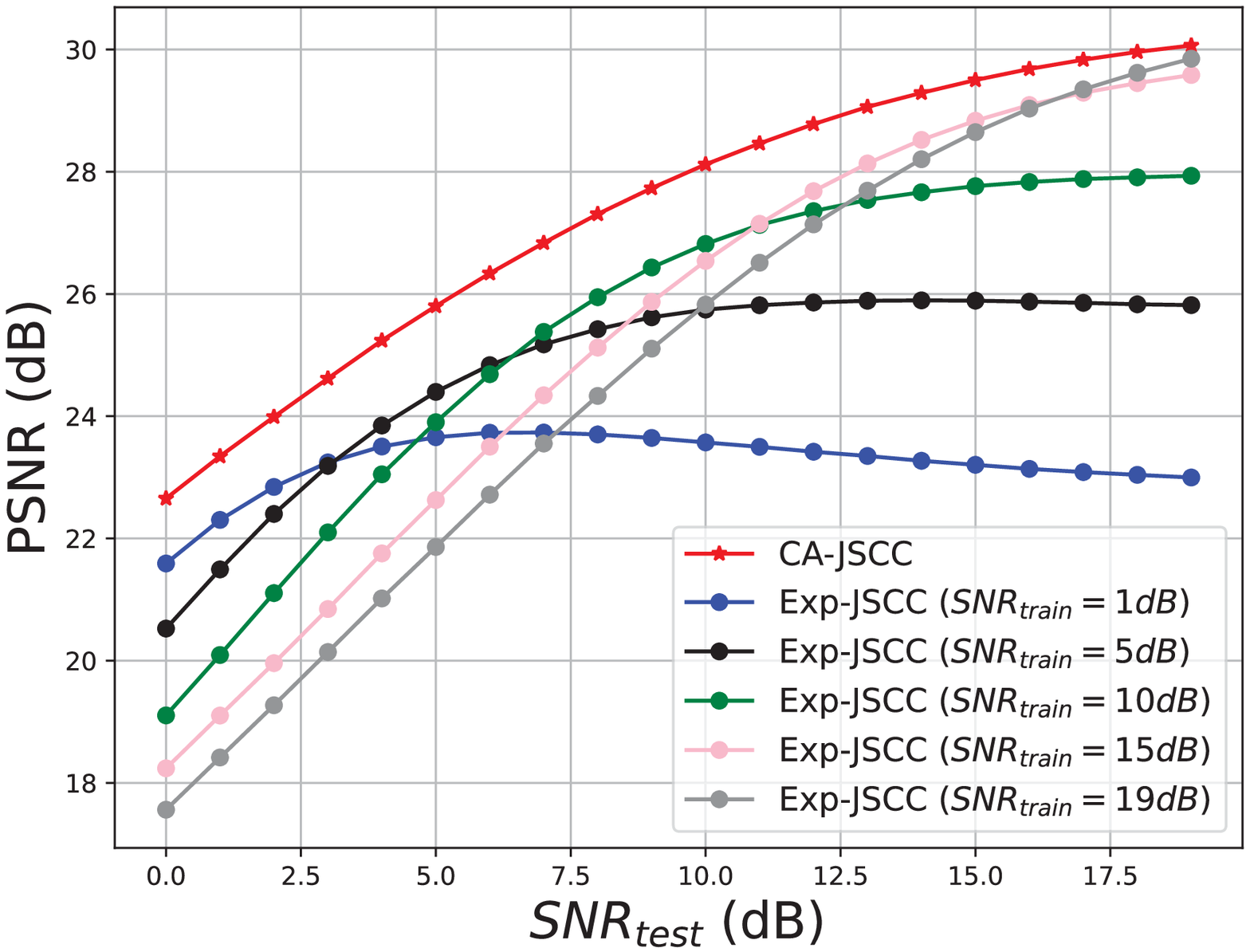}
    \end{minipage}%
    }%
    \hspace{-0.5mm}
    \subfloat[$R=1/3$]{
    \label{attention_JSCC_3} 
    \begin{minipage}[t]{0.32\linewidth}
    \centering
    \includegraphics[scale=0.33]{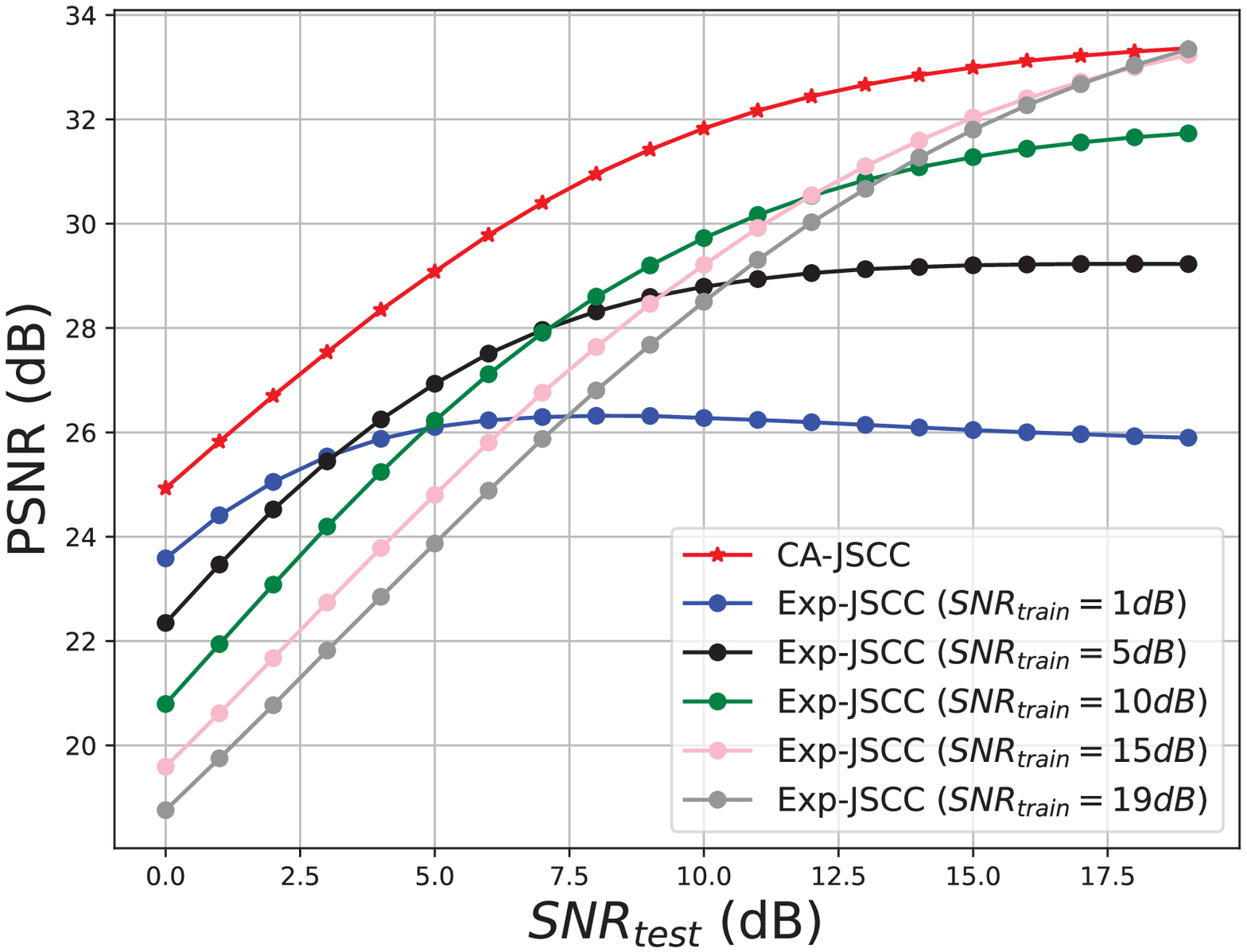}
    \end{minipage}%
    }%
    \centering
    \caption{Performance of our CA-JSCC model comparied with the Exp-JSCC model of different bandwidth ratios.}
    \label{attention_JSCC_perf} 
\end{figure*}

\section{Training and evaluation}
This section presents numerical experiment results to evaluate the performance of our CA-JSCC scheme. The Exp-JSCC scheme introduced in \cite{yang2021ofdm} is the most related work to the current study in the literature. We use Exp-JSCC trained on different channel conditions as a benchmark to compare with the proposed CA-JSCC scheme.

\subsection{Experimental setup}
If not specified otherwise, all experiments were performed on the CIFAR-10 dataset \cite{krizhevsky2009learning} with PyTorch. Models were trained until the performance on a validation set (selected separately from the training dataset) stops improving.
The Adam optimizer is used to perform backpropagation.

We set the number of subcarriers to  $L_{f}=64$. The Zadoff-Chu (ZC) sequence \cite{chu1972polyphase}, denoted by $\bm{Y_p}\in \mathbb{C}^{(2,64)}$, is used as the pilot. The values of channel gains $\{H[k]:k=1,2,...,L_f\}$ are sampled from a complex Gaussian distribution $\mathcal{CN}(0,1)$. Unless specified otherwise, the frequency-domain channel responses in the experiments are estimated by an MMSE estimator. We also sort the channels based on their estimated CSI to make training process easier; that is, we have $|H[1]|^2 \geq \cdots \geq|H[L]|^2$. Following\cite{kurka2020deepjscc}, we define the bandwidth ratio (i.e., bandwidth usage to source symbol ratio) as $R\triangleq \frac{N_s L_f}{c\times h\times w}$, where $N_s L_f$ is the number of symbols transmitted per image. 

\subsection{Channel-gain adaptability}
We first verify the adaptability of the CA-JSCC scheme to channel-gain variations. Specifically, under a fixed bandwidth ratio and a given SNR, we want to see if our dual-attention mechanism can instruct the transmitter to exploit better channels and allocate power to different subcarriers judiciously.

The experimental results are shown in Fig \ref{PA_all_SNR}, where we set the number of OFDM symbols to $N_s=8$ and the bandwidth ratio to $R=1/6$. The Exp-JSCC and CA-JSCC models are both trained at a fixed SNR, and tested at the same SNR. As can be seen, by feeding the estimated CSI to the transmitter, CA-JSCC can exploit the channel-gain information and adaptively allocate power to different subcarriers. We can see a significant gain compared to Exp-JSCC at all SNRs. This can be attributed to the advantage of our method in better exploiting the channel gains and allocating power to different subcarriers.

\subsection{SNR adaptability}
Next, we evaluate the SNR adaptability of our scheme. If not specified otherwise, we train the CA-JSCC model at random SNR values of each training episode chosen uniformly from $[0,20]$ dB and test the well-trained model at different SNRs.

Compared with the CA-JSCC scheme trained at specific SNRs in Fig \ref{PA_all_SNR}, we observe that there is a slight performance degradation when it is trained at random SNR values. We conclude that, while CA-JSCC can learn to adapt to different channel SNRs, this flexibility comes at the expense of some loss in the PSNR (up to $1$dB). However, this CA-JSCC model trained at random SNR values still outperforms the Exp-JSCC models trained at specific SNR values.

We also compare the performance of dual-attention based architecture CA-JSCC with an alternative using only channel-wise attention, called (CA-JSCC-CH), as an ablation study. As shown in Fig \ref{Ab_attention}, for three different bandwidth ratios, CA-JSCC architecture outperforms CA-JSCC-CH at all SNR values, which shows that the \textit{spatial} attention mechanism is essential to achieve the improved performance provided by CA-JSCC. We also observe larger gains by our dual-attention method at higher bandwidth ratios and $\text{SNR}_{\text{test}}$ values, where more spatial information and better CSI adaptability benefit both feature mapping and power allocation. To visualize the power allocation executed by CA-JSCC and Exp-JSCC, we plot the average channel gain and the average power allocated to each subcarrier in Fig. \ref{power_allocation}, where we set SNR$=1$dB and R=$1/6$. The channel gains are ordered in an increasing manner in the plot. Compared with Exp-JSCC, Fig. \ref{power_allocation} shows that CA-JSCC generally allocates more power for the subcarrier with better channel conditions, as one would desire. 
 
It is worth noting that the Exp-JSCC scheme is not SNR-adaptive, which means the training and test SNRs of Exp-JSCC must match to achieve a sound performance, as shown in Fig. \ref{PA_all_SNR}. Fig \ref{attention_JSCC_perf} presents the PSNR versus test SNR results for bandwidth ratios of $R=1/12, 1/6, 1/3$ (we set $L_f=64$ and vary $N_s$ to attain different $R$ values).
As stated above, our CA-JSCC scheme can be trained with random SNRs, and yields a single model for each bandwidth ratio to be verified on a range of test SNRs. The Exp-JSCC scheme, on the other hand, is trained at five different SNRs, yielding five different models under each bandwidth ratio. The Exp-JSCC scheme performs the best when the training and test SNRs match.
However, our CA-JSCC scheme is SNR-adaptive and consistently outperforms Exp-JSCC at all SNRs and bandwidth ratios with a considerable margin.
 
\begin{figure}[t]
     \centering
    \subfloat[]{
    \label{Imagenet}
     \begin{minipage}[t]{1\linewidth}
    \centering
    \includegraphics[scale=0.38]{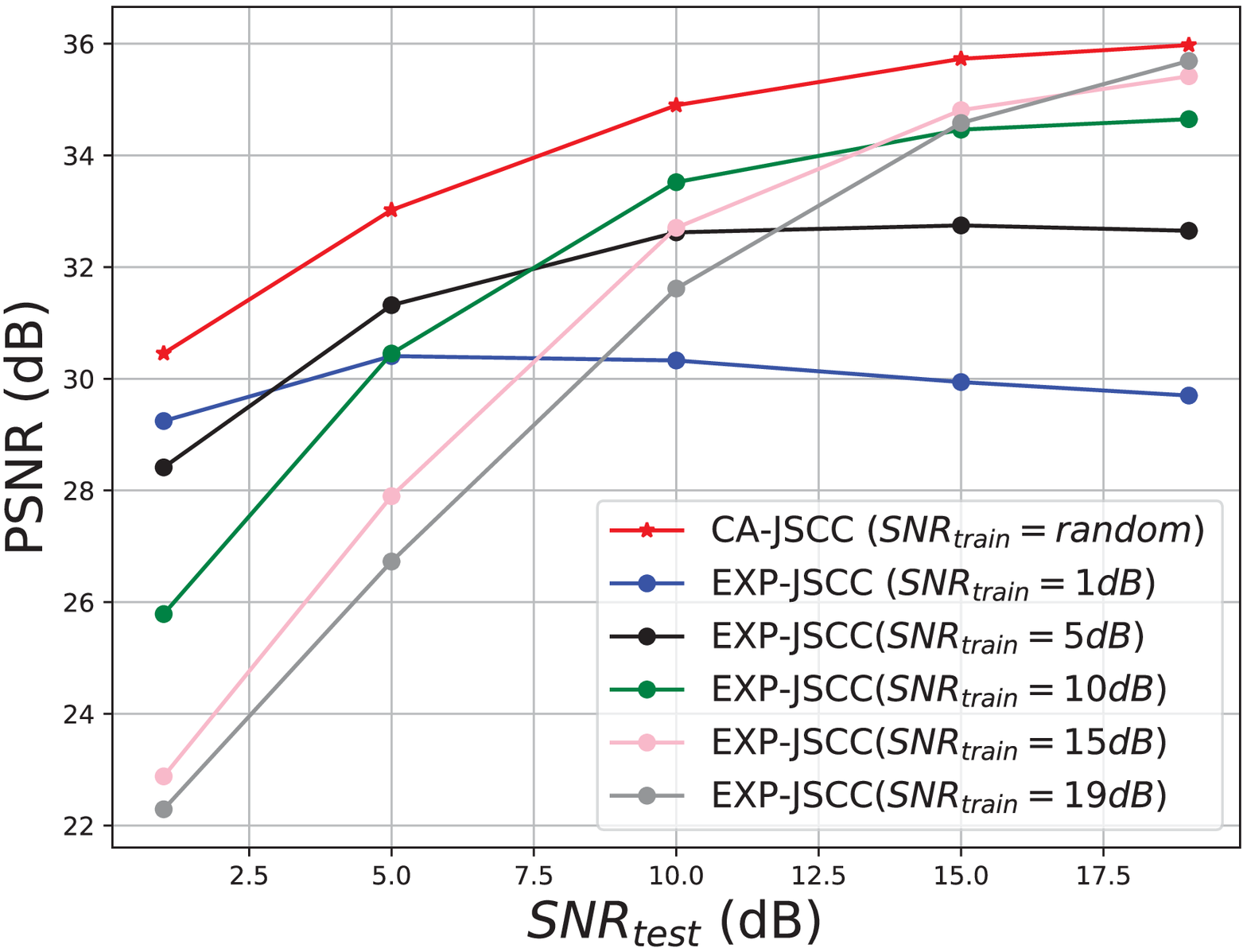}
     \end{minipage}%
     }%
     \hspace{-1.5mm}
     \subfloat[]{
     \label{diff_CSI}
     \begin{minipage}[t]{1\linewidth}
     \centering
    \includegraphics[scale=0.38]{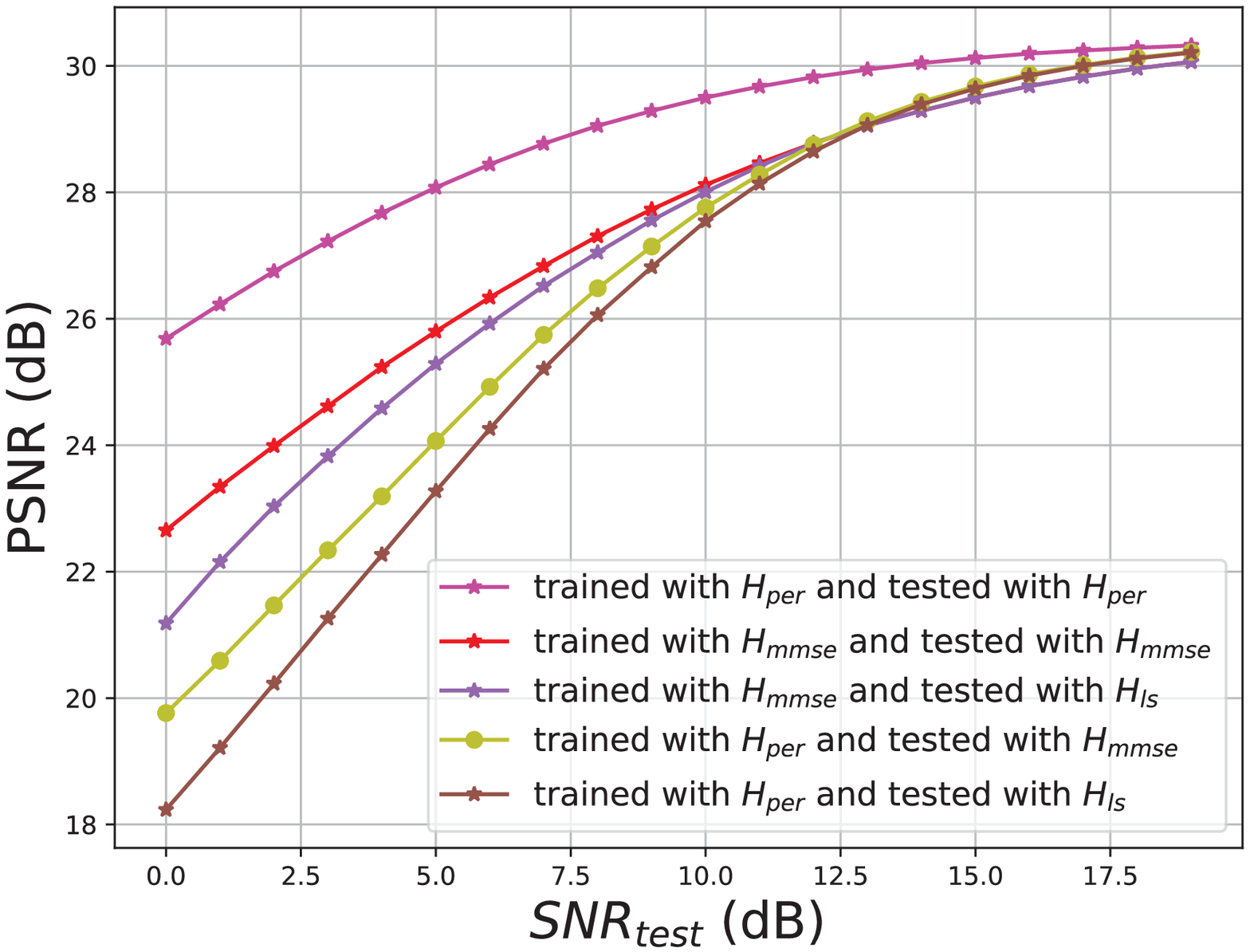}
     \end{minipage}%
     }%
     \caption{(a) Additional experiments with training on the ImageNet dataset tested on the Kodak dataset. (b) Comparison of CA-JSCC with different channel estimation methods.}
 \end{figure}

Additional experiments by training over the ImageNet dataset is shown in Fig. \ref{Imagenet}. We train the models with randomly cropped $64 \times 64$ patches from ImageNet, and evaluate the models on the Kodak dataset. Results show that training on a sufficiently large dataset (ImageNet) can allow our CA-JSCC model to perform well on a never-seen dataset (Kodak). CA-JSCC can still achieve state-of-the-art performance with the additional capability of channel adaptability.
\subsection{Impact of CSI estimation errors}
In the above experiments, we have assumed MMSE estimated CSI. In this subsection, we look into the effect of channel estimation errors on the performance of CA-JSCC. We repeat the experiment in Fig \ref{attention_JSCC_6} with three types of CSI: i) perfect CSI, $H_{per}$; ii) MMSE estimated CSI, $H_{mmse}$; and iii) LS estimated CSI, $H_{ls}$. We remark that $H_{mmse}$ provides a more accurate estimate than $H_{ls}$.

The experimental results are shown in Fig \ref{diff_CSI}, where we train our CA-JSCC models with $H_{mmse}$ and $H_{per}$, respectively, and evaluate these two models with $H_{per}$, $H_{mmse}$ and $H_{ls}$, respectively. As expected, the model trained and tested with $H_{per}$ achieves the best performance. On the other hand, the model trained with perfect CSI is not robust to CSI errors during test time. Its performance gets worse as the quality of channel estimation degrades. Instead, we see that models trained with $H_{mmse}$ perform better, since during training they learn to compensate for CSI estimation errors. We conclude from these results that a more accurate CSI during testing is generally beneficial, and the performance improves if training is done with the same type of CSI.

\section{Conclusion}
We presented the CA-JSCC scheme for wireless image transmission over OFDM channels. CA-JSCC employs a dual-attention mechanism to exploit the available CSI to simultaneously aid the allocation of features and power resources adaptively.
The dual-attention mechanism comprises a \textit{channel-wise} attention module, which learns to map features to subcarriers according to the CSI, and a \textit{spatial} attention module, which learns high-level spatial features, allowing the JSCC encoder to match the most important features with the most reliable subcarriers. Numerical experiments show that our method achieves state-of-the-art performance in various SNR and bandwidth scenarios. Besides, our method simultaneously uses the estimated CSI to adapt to time-varying channel conditions.


\bibliographystyle{IEEEtran}
\bibliography{ref}

\vfill

\end{document}